\documentclass[
aps,pra,
twocolumn,
showpacs,
superscriptaddress,
floatfix,
]{revtex4-1}
\usepackage[dvipdfmx]{graphicx}
\usepackage{dcolumn}
\usepackage{bm}
\usepackage{ulem}
\usepackage{amsmath}
\usepackage{amssymb}
\usepackage{txfonts}
\usepackage{hyperref}
\usepackage{color} 
\hypersetup{
  colorlinks=true,
  linkcolor=[rgb]{0.85,0.00,0.15},
  citecolor=[rgb]{0.0,0.15,0.95},
  urlcolor=[rgb]{0.0,0.15,0.95},
  setpagesize=false
}

\begin{document}

\title{
Topological insulators in ordered double transition metals M$'_2$M$''$C$_2$ 
(M$'$= Mo, W; M$''$= Ti, Zr, Hf) MXenes
}

\author{Mohammad Khazaei
\footnote{E-mail address: khazaei@riken.jp}}
\author{Ahmad Ranjbar}
\affiliation{Computational Materials Science Research Team, RIKEN Advanced Institute for Computational Science (AICS), Kobe, Hyogo 650-0047, Japan}

\author{Masao Arai}
\affiliation{International Center for Materials Nanoarchitectonics, National Institute for Materials Science (NIMS), 1-1 Namiki, Tsukuba 305-0044, Ibaraki, Japan}

\author{Seiji Yunoki}
\affiliation{Computational Materials Science Research Team, RIKEN Advanced Institute for Computational Science (AICS), Kobe, Hyogo 650-0047, Japan}
\affiliation{Computational Condensed Matter Physics Laboratory, RIKEN, Wako, Saitama 351-0198, Japan}
\affiliation{Computational Quantum Matter Research Team, RIKEN Center for Emergent Matter Science (CEMS), Wako, Saitama 351-0198, Japan}

\date{\today}

\begin{abstract}
The family of two-dimensional transition metal carbides, 
so called MXenes, has recently 
found new members with ordered double transition metals M$'_2$M$''$C$_2$,
where M$'$ and M$''$ stand for transition metals. 
Here, using a set of first-principles calculations, we demonstrate 
that some of the newly added members, oxide M$'_2$M$''$C$_2$ (M$'$= Mo, W; M$''$= Ti, Zr, Hf) MXenes, are 
topological insulators. The nontrivial topological states of the predicted MXenes are revealed 
by the $Z_2$ index, which is evaluated from the parities of the occupied bands below the Fermi energy 
at time reversal invariant momenta, and also by the presence of the edge states. 
The predicted M$'_2$M$''$C$_2$O$_2$ MXenes show nontrivial gaps in the range of 
0.041 -- 0.285 eV within the generalized gradient approximation and  
0.119 -- 0.401 eV within the hybrid functional.  
The band 
gaps are induced by the spin-orbit coupling within the degenerate states with $d_{x^2-y^2}$ and $d_{xy}$ 
characters of 
M$'$ and M$''$, while the band inversion occurs at the $\Gamma$ point 
among the degenerate $d_{x^2-y^2}$/$d_{xy}$ orbitals 
and a non-degenerate $d_{3z^2-r^2}$ orbital, which is driven by the hybridization 
of the neighboring orbitals. 
The phonon dispersion calculations find that the predicted topological insulators are structurally stable. 
 The predicted W-based MXenes with large 
band gaps might be suitable candidates for many topological applications at room temperature.
In addition, we study the electronic structures of thicker ordered double transition metals 
M$'_2$M$''_2$C$_3$O$_2$ (M$'$= Mo, W; M$''$= Ti, Zr, Hf) and find that 
they are nontrivial topological semimetals.  
Among the predicted topological insulators and topological semimetals, Mo$_2$TiC$_2$ and 
Mo$_2$Ti$_2$C$_3$ functionalized with mixture of F, O, and OH have already been synthesized, 
and therefore some of the topological materials proposed here can be experimentally accessed. 
\end{abstract}


\maketitle
\section{INTRODUCTION}

Topological insulators (TIs)~\cite{hasan2010,qi2011,yan2012_1,weng2015_1} promise an avenue to realize fascinating applications such as dissipationless 
transport, spintronics, optoelectronics, thermoelectronics, fault-tolerant quantum computing, and efficient 
power transition~\cite{bernevig2006,hsieh2009,chen2009,zhang2009,lin2010,chadov2010,zhang2012,muchler2012_1,muchler2012_2,liu2014}. This is due 
to their unique surface states that are topologically protected and thus robust against non-magnetic impurities 
and disorders. 
The existence of these remarkable electronic states in TIs 
is attributed to the large spin-orbit coupling (SOC) of their heavy elements. TIs can have two- or three-dimensional 
structures. The two-dimensional (2D) TIs possess two wire-like metallic edge states in which electrons propagate 
with opposite spins~\cite{sun2015,zhou2015,tang2014,yan2012_2,ma2015_1,chuang2014,chuang2013,xu2013,ma2016,pham2015,ma2015_2,yang2014,liu2016,wang2015,zhao2016,chou2014,
kou2015_1,kou2015_2,zhang2015_1,song2014,wang2013,wang2014,si2014,sung2016}. 
The three-dimensional TIs have metallic surface states that usually form a single or an odd number of Dirac 
cones at or around the Fermi level~\cite{lin2010,muchler2012_2,sun2015}.

Recently, 2D systems have received a lot of attentions because of their high tunability of charge, spin, and orbital 
states as well as electron confinement by surface functionalization and thickness control. 
Owing to the great advance in experimental techniques in recent years, the number of synthesized 
2D systems has significantly increased~\cite{zhang2015_2}. Among the recent synthesized structures, 2D transition 
metal carbides and nitrides, so called MXenes, have attracted considerable attentions due to their high 
mechanical stability, various elemental compositional and surface functional possibilities, 
and flexible thickness controllability. 
MXenes are a new class of 2D transition metal carbides and nitrides with chemical formula of 
M$_{n+1}$X$_n$ (M= Sc, Ti, V, Cr, Zr, Nb, Mo, Hf, Ta; X= C, N) with $n=1,2,3$ 
that have recently been synthesized through 
hydrofluoric etching~\cite{naguib2011,naguib2012} of layered MAX phase compounds$ - $M$_{n+1}$AX$_n$, where 
A = Al, Si, P, S, Ga, Ge, As, In, and Sn~\cite{khazaei2014_1,khazaei2014_2}. 
During the etching process, the A element is washed out from 
the MAX phase structure and simultaneously the surfaces of the resulting 2D systems are chemically saturated 
with mixture of F, O, and OH~\cite{naguib2011,naguib2012,hope2016,haris2015}. These 2D systems have been named 
MXenes because they originate from the MAX phases by removing A elements and because they are 
structurally analogous to the graphene~\cite{naguib2011,naguib2012}. 
The 2D MXenes such as Ti$_2$C, V$_2$C, Nb$_2$C, Ta$_2$C, Mo$_2$C, Ti$_3$C$_2$, Nb$_4$C$_3$ have 
already been synthesized~\cite{naguib2011,naguib2012,meshkini2015,yang2016_1}. 
MXenes are known to have or predicted to have electronic, magnetic, and energy harvesting 
applications~\cite{khazaei2013,khazaei2016, khazaei2014_3, khazaei2015, si2015, bai2016,ling2016,
yang2016_2,hong2016,zha2016_1,he2016,gao2016,xie2014,yorulmaz2016,zha2016_2}.

Theoretically, it has been shown that some of the MXenes possess Dirac band dispersions crossing 
at the Fermi level, owning to the honeycomb like structure~\cite{fashandi2015}. 
Therefore, they are highly suspected for being TIs if heavy transition metal elements are involved. 
Indeed, we have previously shown that 
among the members 
of functionalized M$_2$X MXenes, Mo$_2$CO$_2$ and W$_2$CO$_2$ are TIs~\cite{weng2015_2}. 
The family of MXenes has lately been expanded to the ordered double transition metals carbides 
M$'_2$M$''$C$_2$ and M$'_2$M$''_2$C$_3$, where M$'$ and M$''$ stands for transition 
metals~\cite{anasori2016_1,anasori2016_2,anasori2015}. 
This breakthrough experiment is a major success because previously the MXenes with different transition 
metals such as TiNbC, (Ti$_{0.5}$V$_{0.5}$)$_2$C, (V$_{0.5}$Cr$_{0.5}$)$_3$C$_2$, Ti$_3$CN, and 
(Nb$_{0.8}$Ti$_{0.2}$)$_4$C could only be synthesized in the alloy forms~\cite{naguib2011,
naguib2012,mashtalir2013, lukatskaya2013}, but not the ordered ones. Here, using first-principles 
electronic structure calculations, we demonstrate that in this newly discovered MXenes, 
oxide M$'_2$M$''$C$_2$ (M$'$ = Mo, W; M$''$ = Ti, Zr, Hf) are TIs with band gaps as large as 
0.285 eV within the generalized gradient approximation 
0.401 eV within the hybrid functional)  
and oxide M$'_2$M$''_2$C$_3$ are nontrivial topological semimetals. 
Because Mo$_2$TiC$_2$ and Mo$_2$Ti$_2$C$_3$ MXenes with mixture of 
F, OH, and O surface chemical groups has already been synthesized~\cite{anasori2016_1,anasori2016_2}, 
some of the proposed oxide M$'_2$M$''$C$_2$ and M$'_2$M$''_2$C$_3$ will be experimentally accessed 
in the near future.

\section{METHOD OF CALCULATIONS}

The structural optimizations and electronic structure calculations are performed in the context of density 
functional theory as implemented in VASP code~\cite{vasp1996}. 
Exchange-correlation energies are taken into account by the generalized gradient 
approximation (GGA) using Perdew-Burke-Ernzerhof functional~\cite{pbe1996}. 
The wave functions are constructed using projected augmented wave approach with plane wave cutoff 
energy of 520 eV. The effect of spin-orbit coupling (SOC) is included self-consistently in the electronic 
structure calculations. The atomic positions and cell parameters are fully optimized using 
conjugate gradient method without imposing any symmetry. After the optimization process, the maximum 
residual force on each atom is less than 0.001 eV/\AA. The total energies are converged within 
10$^{-6}$ eV/cell. A large vacuum space of 50~\AA~ 
is set along the $c$ axis, 
the direction perpendicular to the surface, to avoid any interaction between the layer and its periodic images. 
The Brillouin zone integration is sampled using a set of 12$\times$12$\times$1 Monkhorst-Pack 
{\bf k} points~\cite{monkhorst1976}. 
Since the GGA often underestimates the band gap, we also perform the hybrid functional 
[Heyd-Scuseria-Ernzerhof (HSE06)] calculations~\cite{heyd2003,heyd2006} 
with 20 {\bf k} points along each path section to check the band topology. 
 
The phonon dispersions are obtained using the density functional perturbation theory in the same level of 
approximation as described above using Quantum Espresso code~\cite{baroni2001,giannozzi2009}. 
The edge states are calculated using 
a nonuniform tight-binding Green's function method with maximally localized Wannier function 
basis sets~\cite{marzari1997,souza2001}
generated by OpenMX code~\cite{openmx}.

\section{RESULTS AND DISCUSSION}

\subsection{Atomic structure}\label{sec:structure}

\begin{figure}[htbp]
\begin{center}
\includegraphics[width=7.0cm]{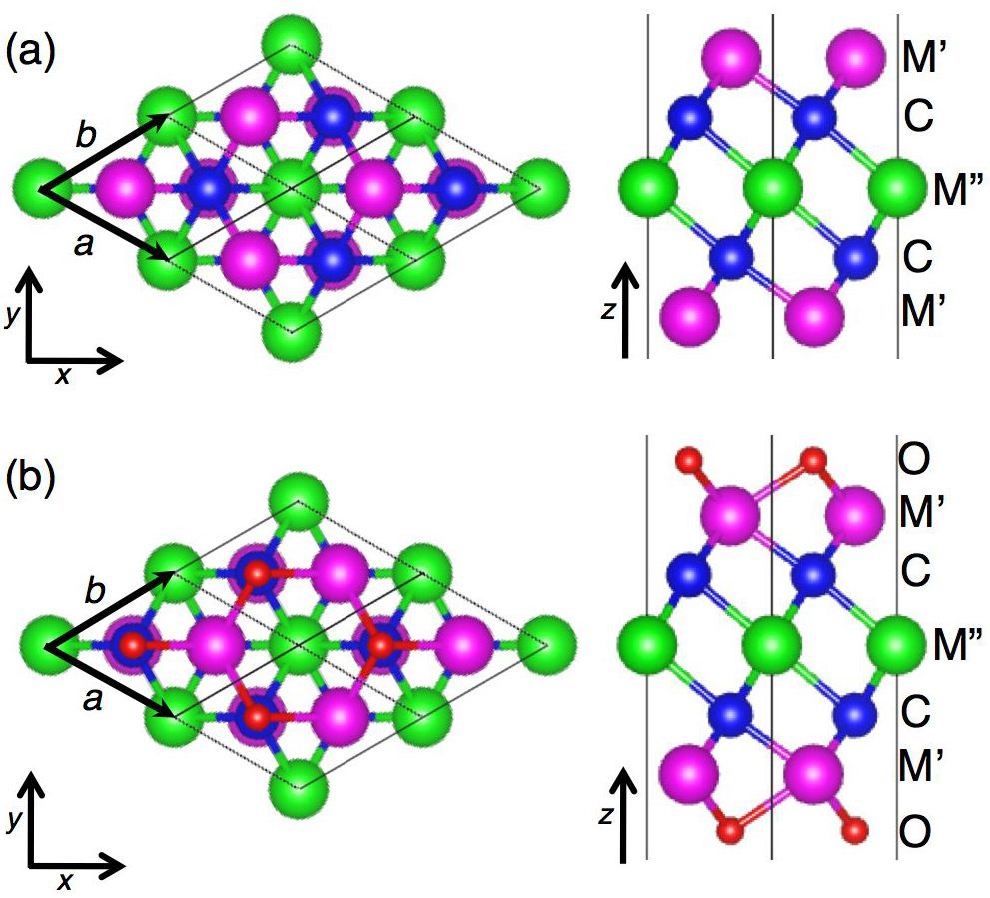}
\end{center}
\caption{(Color online)
Top (left panels) and side (right panels) views of two-dimensional (a) M$'_2$M$''$C$_2$ 
and (b) M$'_2$M$''$C$_2$O$_2$ MXenes. The unit vectors in $ab$ plane are denoted by black arrows.
}
\label{fig:structure}
\end{figure}

As shown in Fig.~\ref{fig:structure}(a), 
M$'_2$M$''$C$_2$ forms a hexagonal lattice 
with $P\bar{3}m1$ group symmetry and is made of five atomic 
layers of M$'$-C-M$''$-C-M$'$ where M$'$ transition metals form the outer surfaces and 
M$''$ transition metals form the central layer. The carbon 
atoms are sandwiched between the M$'$ and M$''$ transition metal layers, and 
each carbon atom is located at the center of an octahedral cage formed by M$'$ and M$''$ 
transition metals. 
Theoretically, it has been shown that the ordered double transition metals MXenes is structurally 
stable, and indeed one of them, Mo$_2$TiC$_2$, has already been synthesized 
experimentally \cite{anasori2016_1,anasori2016_2}.

It is very difficult to synthesize MXenes with pure surfaces from MAX phases by the etching process  
and usually the surfaces of MXenes are saturated with the mixture of chemical groups that depend on 
the type of applied chemical solution. 
For instance, when hydrofluoric acid is used, the surfaces of MXenes are saturated with the mixture of 
F, O, and OH~\cite{naguib2011,naguib2012}. 
This is due to the high chemical reactivity of transition metals on the exposed surfaces of MXenes. 
It has been shown theoretically that the chemical groups such as F, O, and OH form strong bonding with 
the transition metals and thus the surfaces of MXenes can be fully saturated with them at a proper 
chemical potential~\cite{khazaei2013,ashton2016}. 
Pristine M$'_2$M$''$C$_2$ (M$'$= Mo, W; M$''$= Ti, Zr, Hf) MXenes are metallic~\cite{anasori2016_1}. 
However, upon proper surface functionalization, some of the MXenes become semiconducting~\cite{khazaei2013,anasori2016_2}.
Here, we shall focus on M$'_2$M$''$C$_2$ (M$'$= Mo, W; M$''$= Ti, Zr, Hf) MXenes functionalized with oxygen
to investigate the electronic properties in detail. 
Before studying the electronic structures, the structural properties of M$'_2$M$''$C$_2$O$_2$ are examined.

Two oxygen atoms per unit cell are required to fully functionalize the surfaces of MXenes. 
Generally, the functional groups can be adsorbed on top of the transition metals or on top of hollow sites of the 
surfaces. 
However, the previous calculations find that the configurations with chemical groups adsorbed on top of 
the transition metals sites are energetically unfavorable and these chemical groups eventually move to 
top of the hollow sites after the structural optimization~\cite{khazaei2013,khazaei2014_3}. 
Therefore, the chemical groups are favorably adsorbed on top of the hollow sites formed by the surface 
transition metals. 

The surfaces of MXenes include two types of hollow sites, which are named as A and B here. 
At the B-type (A-type) hollow site, there is (is not) a carbon atom under the hollow. Therefore, 
depending on the relative positions of the attached oxygen groups at the hollow sites of the two surfaces, 
three different model configurations are possible for the chemical terminations of M$'_2$M$''$C$_2$: 
AA, AB, and BB models. 
In AA (BB) model, the two oxygen atoms are adsorbed on top of the A-type (B-type) hollow sites, 
while in AB model, one of the oxygen atoms is adsorbed on top of the A-type hollow site and the other one 
is adsorbed on top of the B-type hollow.

In order to find the most stable model structures, we fully optimized the structures of the above three models 
for each M$'_2$M$''$C$_2$O$_2$ system. The total energies of the optimized models are summarized 
in Table~\ref{tab:TAB1}. 
Since the total energy is lowest for the BB systems, the BB model is considered to be the most 
appropriate for the M$'_2$M$''$C$_2$O$_2$ (M$'$ = Mo, W; M$''$ = Ti, Zr, Hf) MXenes. 
The most stable BB type structure for the M$'_2$M$''$C$_2$O$_2$ MXenes is shown in 
Fig.~\ref{fig:structure}(b) and the detailed crystal parameters are found in Supplemental 
Material~\cite{supplementary}.

\begin{table}
\caption{
The total energy (in eV) per unit cell of the optimized AA, AB, and BB models for 
M$'_2$M$''$C$_2$O$_2$ (M$'$ = Mo, W; M$''$ = Ti, Zr, Hf).
}
\begin{center}
\begin{tabular}{cccc}
\hline
\hline
  & AA  &  AB   &  BB \\
\hline 

Mo$_2$TiC$_2$O$_2$ &  $-64.057$ & $-64.628$ & $-65.238$ \\
Mo$_2$ZrC$_2$O$_2$ & $-64.436$ & $-64.868$ & $-65.388$    \\
Mo$_2$HfC$_2$O$_2$ & $-66.278$ & $-66.765$ & $-67.328$ \\
W$_2$TiC$_2$O$_2$   & $-68.336$ & $-68.911$ &  $-69.588$ \\
W$_2$ZrC$_2$O$_2$  & $-68.769$ & $-69.187$ &  $-69.747$ \\
W$_2$HfC$_2$O$_2$  & $-70.571$ & $-71.075$ &  $-71.694$ \\
 \hline
\hline
\end{tabular}
\end{center}
\label{tab:TAB1}
\end{table}

\begin{figure}[htbp]
\begin{center}
\includegraphics[width=7.0cm]{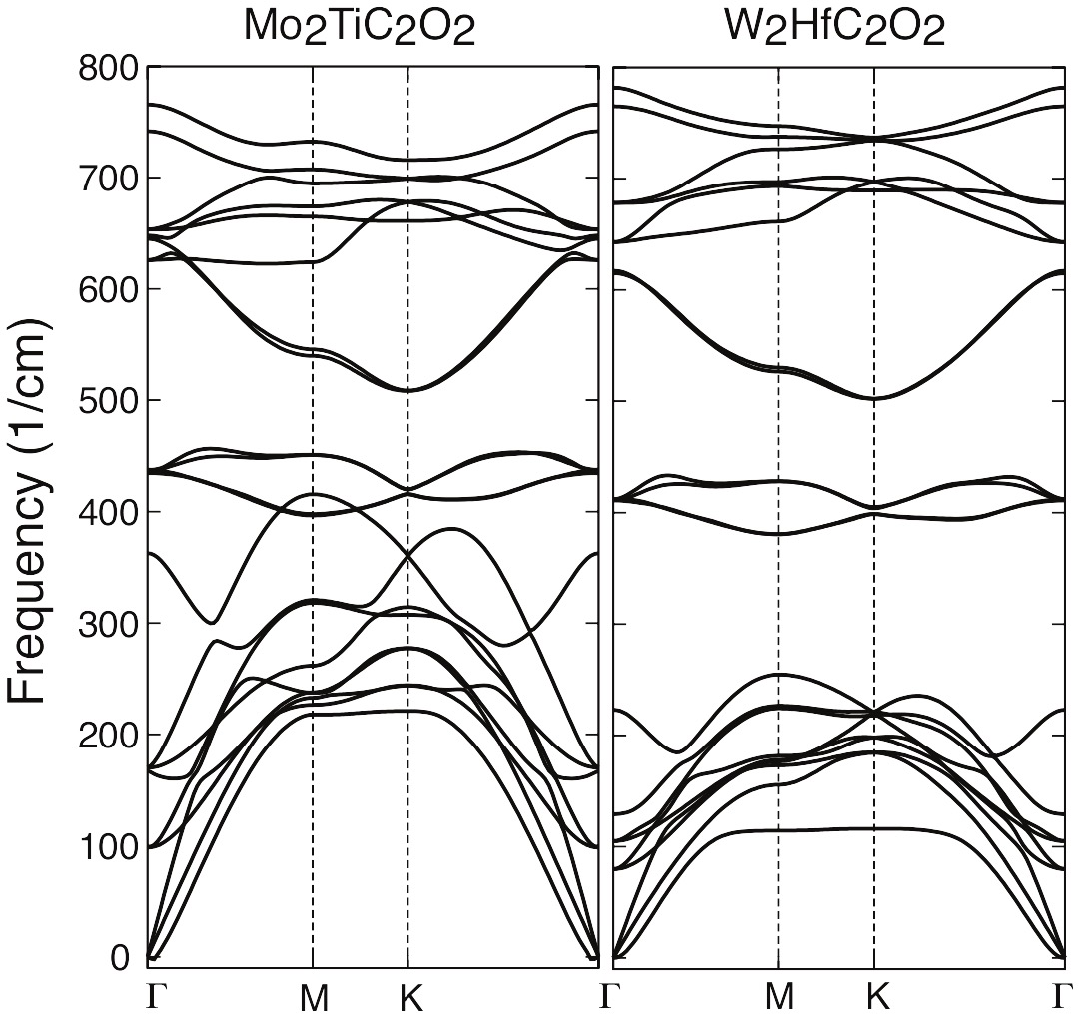}
\end{center}
\caption{
Phonon dispersions for Mo$_2$TiC$_2$O$_2$ and W$_2$HfC$_2$O$_2$. 
}
\label{fig:FIG2}
\end{figure}

In order to ensure that all atoms in the predicted M$'_2$M$''$C$_2$O$_2$ structures are in their 
equilibrium positions, we also calculated a set of phonon dispersions. Typical examples of the phonon 
dispersions for Mo$_2$TiC$_2$O$_2$ and W$_2$HfC$_2$O$_2$ are shown in Fig.~\ref{fig:FIG2}, 
the results for others are included into Supplemental Material~\cite{supplementary}. 
As shown in Fig.~\ref{fig:FIG2}, we find that all phonon modes are positive, indicating that the predicted 
structures are dynamically stable. The highest phonon frequency in M$'_2$M$''$C$_2$O$_2$ 
MXenes is around 780 cm$^{-1}$, which is higher than that for MoS$_2$ 475 cm$^{-1}$~\cite{zhang2015_3}, 
but lower than that for graphene 1600 cm$^{-1}$~\cite{mafra2014}. 
This indicates the M$'_2$M$''$C$_2$O$_2$ MXenes have higher (lower) stability than MoS$_2$ (graphene). 
The electronic structure analyses given in the following are based on the most stable structures obtained here in this 
section. 

\subsection{Electronic structure}

\begin{figure*}[htbp]
\begin{center}
\includegraphics[width=14.0cm]{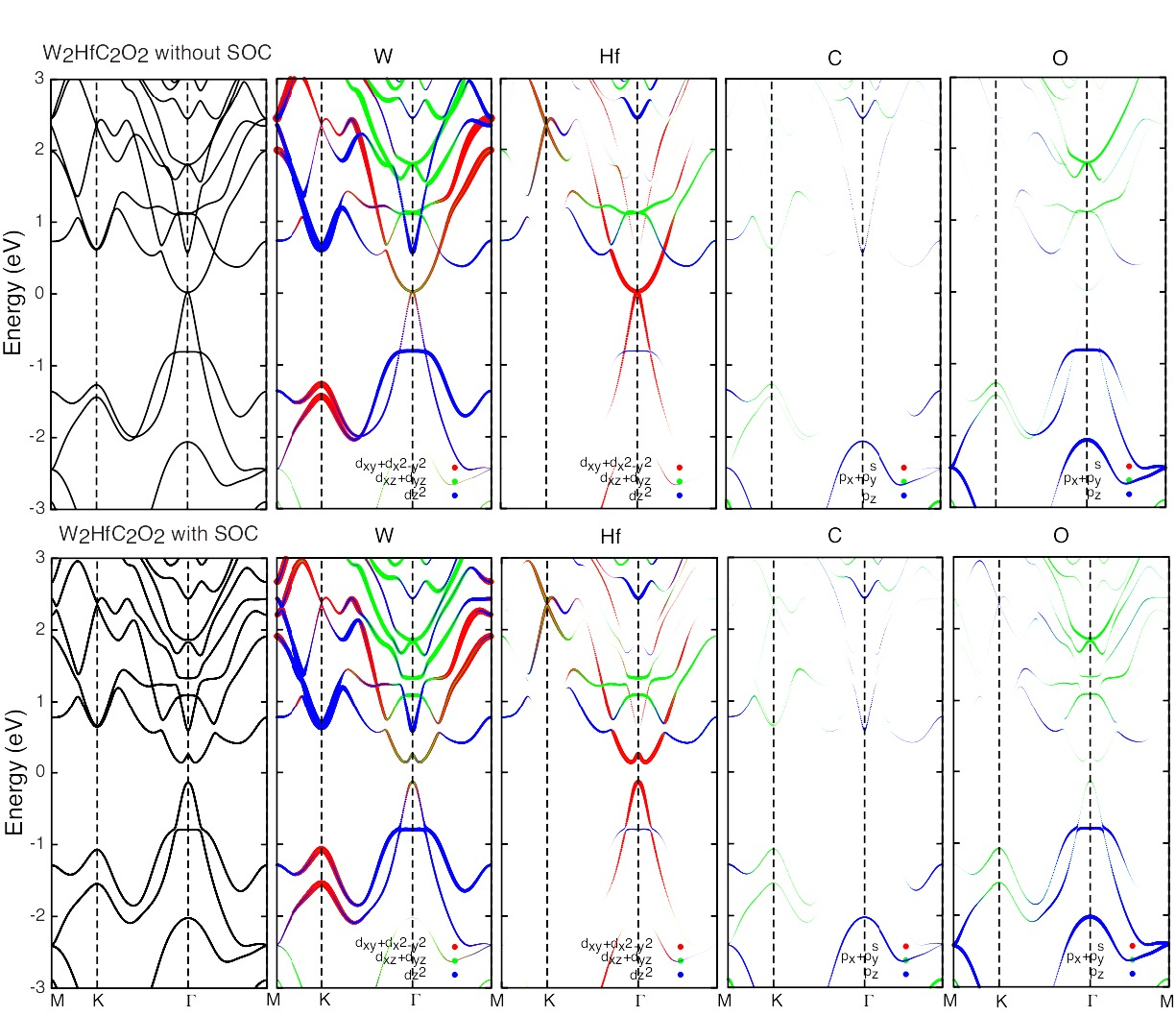}
\end{center}
\caption{(Color online)
Electronic band structures for W$_2$HfC$_2$O$_2$ without (top panels) and with the SOC (bottom panels). 
For detailed analysis, the band structures projected onto each constituent element with different orbital symmetries 
(indicated in the figures) are also shown. 
The Fermi energy is located at zero energy. 
}
\label{fig:proj_dos}
\end{figure*}

Since the band structures are qualitatively the same among all the M$'_2$M$''$C$_2$O$_2$ MXenes studied here, 
we show in Fig.~\ref{fig:proj_dos} the results for W$_2$HfC$_2$O$_2$ with and without the SOC. 
The band structures for other MXenes are given in Supplemental Material~\cite{supplementary}. 
As shown in Fig.~\ref{fig:proj_dos}, when the SOC is not considered, the system is semimetallic because 
the topmost valence band and the lowest conduction band touch only at the $\Gamma$ point, around which the 
valence and conduction bands are both parabolic. 
In order to better analyze the band structures, the projected band structures onto each constituent element with 
different orbital symmetries are also shown in Fig.~\ref{fig:proj_dos}. 
We can easily find in Fig.~\ref{fig:proj_dos} that the bands near the Femi energy originates from $d$ orbitals of 
transition metals M$'$ and M$''$. 
Because of the hexagonal symmetry of the crystal structure, the $d$ orbitals are categorized into three groups, 
$(d_{xy},d_{x^2-y^2})$, $(d_{xz},d_{yz})$, and $d_{3z^2-r^2}$, although the former two groups belong to the 
same irreducible representation for $D_{3d}$ symmetry. 
Similarly, the $p$ orbitals are divided into two groups, $(p_x,p_y)$ and $p_z$.
As shown in Fig.~\ref{fig:proj_dos}, indeed, the topmost valence band and the lowest conduction band 
at the $\Gamma$ point around the Fermi energy are dominated by $d_{xy}$ and $d_{x^2-y^2}$ orbitals 
of transition metals M$'$ and M$''$.

Often, 2D materials composed of heavy elements with touching the valence and conduction bands at the 
Fermi energy are suspected for being TIs. This motivates us to examine the possible nontrivial band topology 
of M$'_2$M$''$C$_2$O$_2$ MXenes. 
As shown in Fig.~\ref{fig:proj_dos}, upon considering the SOC, the degeneracy of the topmost valence band and 
the lowest conduction band at the Fermi energy is lifted (except for the Kramers degeneracy) and the band 
gap is open. 
Consequently, the M$'_2$M$''$C$_2$O$_2$ MXenes (M$'$= Mo, W; M$''$ = Ti, Zr, Hf) turn into semiconductor 
with indirect band gaps, as summarized in Table~\ref{tab:gap}. 
It is clearly observed in Table~\ref{tab:gap} that 
the induced band gap is larger as the SOC (i.e, mass) of M$'$ and/or M$''$ is larger. 
We find that the band gap can be as large (small) as 0.285 eV (0.041 eV) for W$_2$HfC$_2$O$_2$ 
(Mo$_2$TiC$_2$O$_2$).

We further explore the effect of SOC on the band gap by turning on the SOC for a particular element, 
either M$'$ or M$''$ in M$'_2$M$''$C$_2$O$_2$. 
This can be done using Quantum Espresso~\cite{baroni2001,giannozzi2009}. 
As expected, the results of the band gaps calculated using VASP and Quantum Espresso are almost 
the same when the SOC is on for all the elements (see Table~\ref{tab:gap}). 
When the SOC is switched on only for M$'$ or M$''$ element, the degeneracy of the topmost valence band 
and the lowest conduction band at the $\Gamma$ point is lifted and the band gap is open, 
as summarized in Table~\ref{tab:gap} (and also see Supplemental Material~\cite{supplementary}), 
for all the M$'_2$M$''$C$_2$O$_2$ MXenes studied here, 
implying that the strength of the SOC for either  M$'$ or M$''$ is enough to open a gap in these systems. 
The degeneracy at the $\Gamma$ point should be lifted because there is no four-dimensional irreducible 
representation for $D_{3d}$ double group. 
However, notice that these band gaps are always smaller than those evaluated with the SOC on for both 
M$'$ and M$"$ elements. 
Comparing the band gaps for the different systems in Table~\ref{tab:gap}, we find that 
the SOC for each transition metal seems to contribute separately to opening the gap as large as 
$\sim$0.026, $\sim$0.123, $\sim$0.01, $\sim$0.04, and $\sim$0.14 eV 
for Mo, W, Ti, Zr, and Hf, respectively. 
It is well known that the GGA calculations underestimate the band gaps. Hence, 
in order to better estimate the band gap as well as the band topology, we examine
the band structures using the hybrid functionals (HSE06). 
As shown in the Supplemental Material~\cite{supplementary}, the similar band topologies are found 
through the hybrid calculations and the band gaps are estimated to be in the range of 0.119 -- 0.401 eV 
(see Table~\ref{tab:gap}).
To ensure that these insulators are topologically nontrivial, next we shall calculate the $Z_2$ topological 
invariant.

Since all the M$'_2$M$''$C$_2$O$_2$ MXenes studied here have the inversion symmetry, 
their Z$_2$ topological invariant can be simply calculated from the parity of their valence band wave 
functions at the time reversal invariant momentum (TRIM) points of the Brillouin zone~\cite{fu2007_1,fu2007_2}. 
More precisely, the $Z_2$ index $\nu$ is evaluated as $(-1)^\nu=\Pi_{i=1}^4\delta(k_i)$, where 
$\delta(k_i)=\Pi_{n=1}^{N}\zeta_n^i$, $\zeta_n^i$ $(=\pm1)$ is the parity of the $n$th valence band at 
the $i$th TRIM $k_i$, and $N$ is the total number of the occupied valence bands~\cite{fu2007_1,fu2007_2}. 
The trivial and nontrivial topological phases are characterized by $\nu$ = 0 and 1, respectively. 
Because the crystal structure of the M$'_2$M$''$C$_2$O$_2$ MXenes is hexagonal (see Fig.~\ref{fig:structure}),  
the TRIM points are at $\Gamma$ point: ${\bf k}=(0,0)$, $M_1$ point: ${\bf k}=(0, 0.5)$, 
$M_2$ point: ${\bf k}=(0.5, 0)$, and $M_3$ point: ${\bf k}=(0.5, 0.5)$. 
Because of the hexagonal symmetry, the parities at the $M_1$, $M_2$, and $M_3$ points become equivalent, 
and commonly are identified as M. Thus the $Z_2$ index can be simply obtained from $(-1)^\nu=\delta^3(M)\delta(\Gamma)$. 
From the parity analysis of the occupied bands at the TRIM points, we find that $\nu$ = 1 for all the systems 
studied here and therefore M$'_2$M$''$C$_2$O$_2$ (M$'$= Mo, W; M$''$= Ti, Zr, Hf) MXenes are TIs.

\begin{figure}[htbp]
\begin{center}
\includegraphics[width=8.5cm]{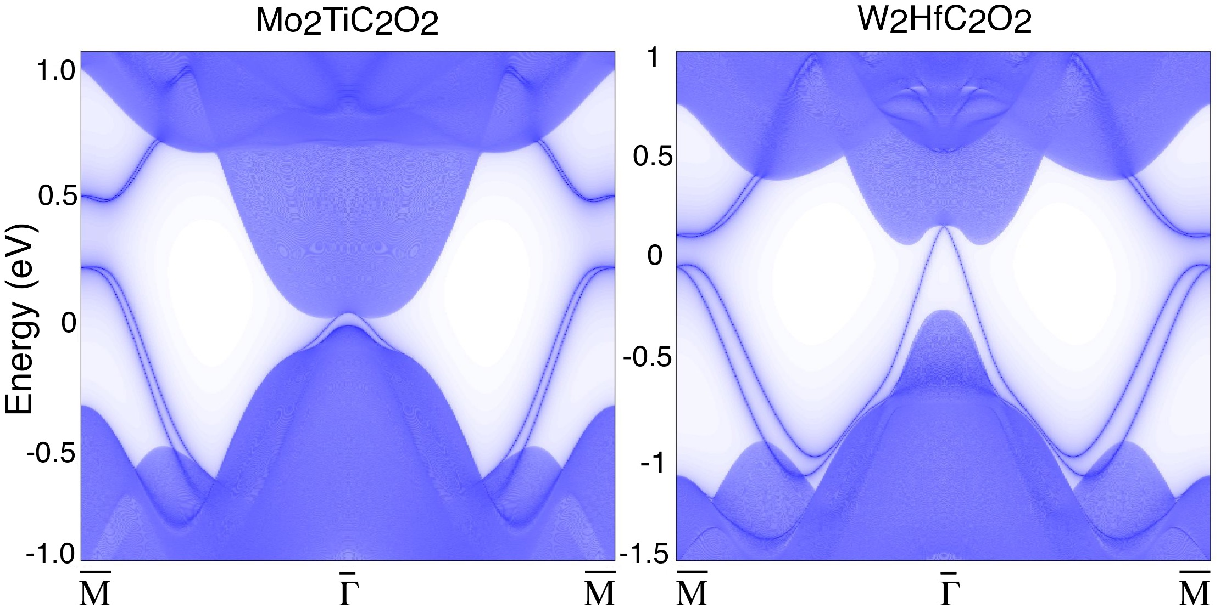}
\end{center}
\caption{(Color online)
Edge band structures for Mo$_2$TiC$_2$O$_2$ and W$_2$HfC$_2$O$_2$. 
The Fermi energy is located at zero energy. 
}
\label{fig:edge}
\end{figure}

One of the remarkable characteristics of the TI is the presence of an odd number of 
topologically protected conducting edge states that cross the Fermi energy. 
In order to further confirm the nontrivial band topology, 
we also calculate the electronic band structures for the nanoribbon structures of Mo$_2$TiC$_2$O$_2$ and 
W$_2$HfC$_2$O$_2$ with symmetric edges 
by using the effective tight binding Hamiltonian, which are constructed based on the maximally localized 
Wannier functions. Since the electronic bands near the Fermi energy are mainly composed of 
$d_{x^2-y^2}$, $d_{xy}$, and $d_{3z^2-y^2}$ orbitals of transition metals M$'$ and M$''$ (see Fig.~\ref{fig:proj_dos}), 
the minimal tight binding Hamiltonian can be constructed using these orbitals. 
Figure~\ref{fig:edge} shows the results of the electronic band structures, which clearly displays that 
the edge bands cross the Fermi energy three times along the $\bar\Gamma$-$\bar M$ points 
for both systems.

\begin{table}
\caption{
Band gaps (in eV) for M$'_2$M$''$C$_2$O$_2$ (M$'$ = Mo, W; M$''$ = Ti, Zr, Hf) obtained using 
VASP and Quantum Espresso (QE)  
with the SOC on for all elements (second column). The results with the SOC on only for M$'$ (M$''$) 
element are also provided in the third (fourth) column, 
obtained using Quantum Espresso.  The band gaps in the second, third, and fourth columns 
are obtained by the GGA.
For comparison, band gaps obtained by the hybrid functional (HSE06) 
calculations using VASP are also provided in the fifth column.
}
\begin{center}
\begin{tabular}{ccccc}
\hline
\hline
           &   SOC is on           &    SOC is        &   SOC is       &  SOC is on \\
           &   for all elements   &   on for       &   on for   &    for all elements \\
           & (VASP/QE)        &   only M$'$     &   only M$''$         & with HSE06 \\
\hline 
Mo$_2$TiC$_2$O$_2$ &  0.041/0.036 & 0.027  &  0.009  &  0.119 \\
Mo$_2$ZrC$_2$O$_2$ &  0.069/0.065 & 0.026 & 0.039   & 0.125 \\
Mo$_2$HfC$_2$O$_2$ & 0.153/0.151 & 0.027 & 0.120   & 0.238 \\
W$_2$TiC$_2$O$_2$   & 0.136/0.135 &  0.121 &  0.010  & 0.290 \\
W$_2$ZrC$_2$O$_2$  & 0.170/0.166 &  0.123 &  0.040 & 0.280 \\
W$_2$HfC$_2$O$_2$  & 0.285/0.285 &  0.135 & 0.140  & 0.409 \\
 \hline
\hline
\end{tabular}
\end{center}
\label{tab:gap}
\end{table}

\subsection{Discussion}

Let us finally examine the origin of the topological insulating behavior found in the 
M$'_2$M$''$C$_2$O$_2$ MXenes.  
As shown in Fig.~\ref{fig:proj_dos}, the SOC is essential to open a finite band gap but does not 
induce the band inversion that is necessary to change the band topology, implying that the 
band inversion has already occurred in the 
M$'_2$M$''$C$_2$O$_2$ MXenes before introducing the SOC. 
This is thus different from most of topological insulators for which the SOC is required to induce 
the band inversion~\cite{chadov2010,lin2010,zhang2012,yan2012_2,yang2014}, although some exceptions have 
been reported~ \cite{zhou2015,ma2016,liu2016,weng2015_2}.

\begin{figure*}[htbp]
\begin{center}
\includegraphics[width=18cm]{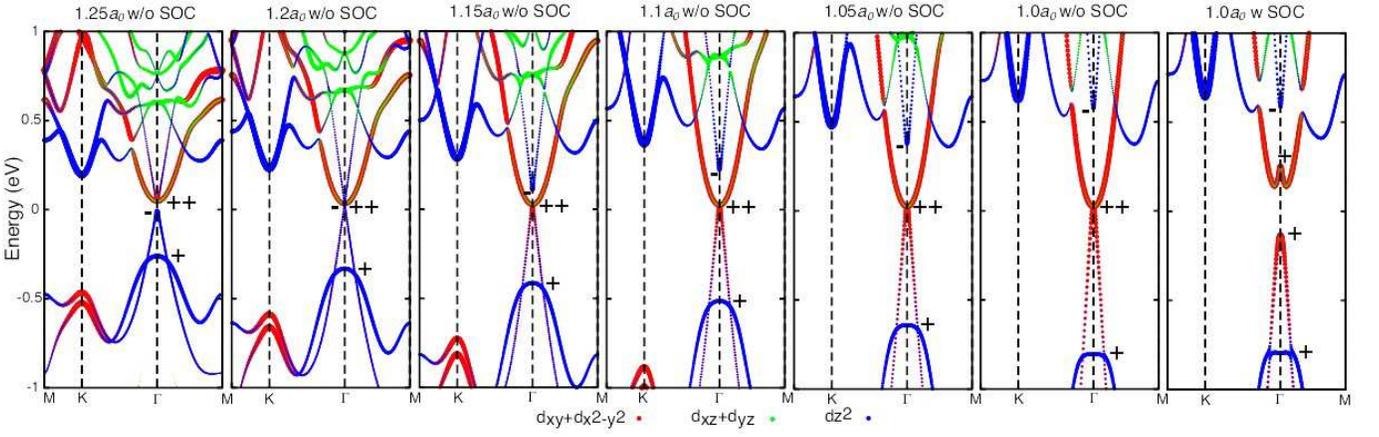}
\end{center}
\caption{(Color online)
Band structure evolution of W$_2$HfC$_2$O$_2$ projected onto 
$d$ orbitals of W and Hf atoms as the lattice constant is expanded uniformly from the optimized one ($a_0$) 
without (w/o) including the SOC. For comparison, the results for the optimized structure with the SOC is also 
shown. Fermi energy is at zero.}
\label{fig:evolution}
\end{figure*}

To understand the band structures in more details, we investigate the evolution of the band structures 
without the SOC as the lattice constant is uniformly expanded from the optimized one that is obtained 
in Sec.~\ref{sec:structure}. Note that the structure is not optimized when the lattice constant is expanded  
because this is much easier to analyze the evolution of the band structures. 
As shown in Fig.~\ref{fig:evolution}, the band structures around the Fermi energy 
are qualitatively the same until the lattice constant is expanded slightly below $20$\% (1.2$a_0$), where 
the topmost valence band and the lowest conduction band remain touched (i.e, degenerate) 
at the $\Gamma$ point and they are essentially composed of $d_{xy}$ and $d_{x^2-y^2}$ orbitals of W and Hf.  
However, when the lattice constant is expanded more than $\sim20$\%, the band gap opens and 
the band characters around the Fermi energy change qualitatively at the $\Gamma$ point, 
where the topmost valence band is 
composed of $d_{3z^2-r^2}$ orbital, while the lowest conduction bands are doubly degenerate and consist of 
$d_{xy}$ and $d_{x^2-y^2}$ orbitals. 
More interestingly, we find that the band inversion occurs 
at the $\Gamma$ point concomitantly when the band gap opens. 
We should note that neither the band inversion nor the gap closing occurs at other momenta except for 
the $\Gamma$ point.

\begin{figure}[htbp]
\begin{center}
\includegraphics[width=8.05cm]{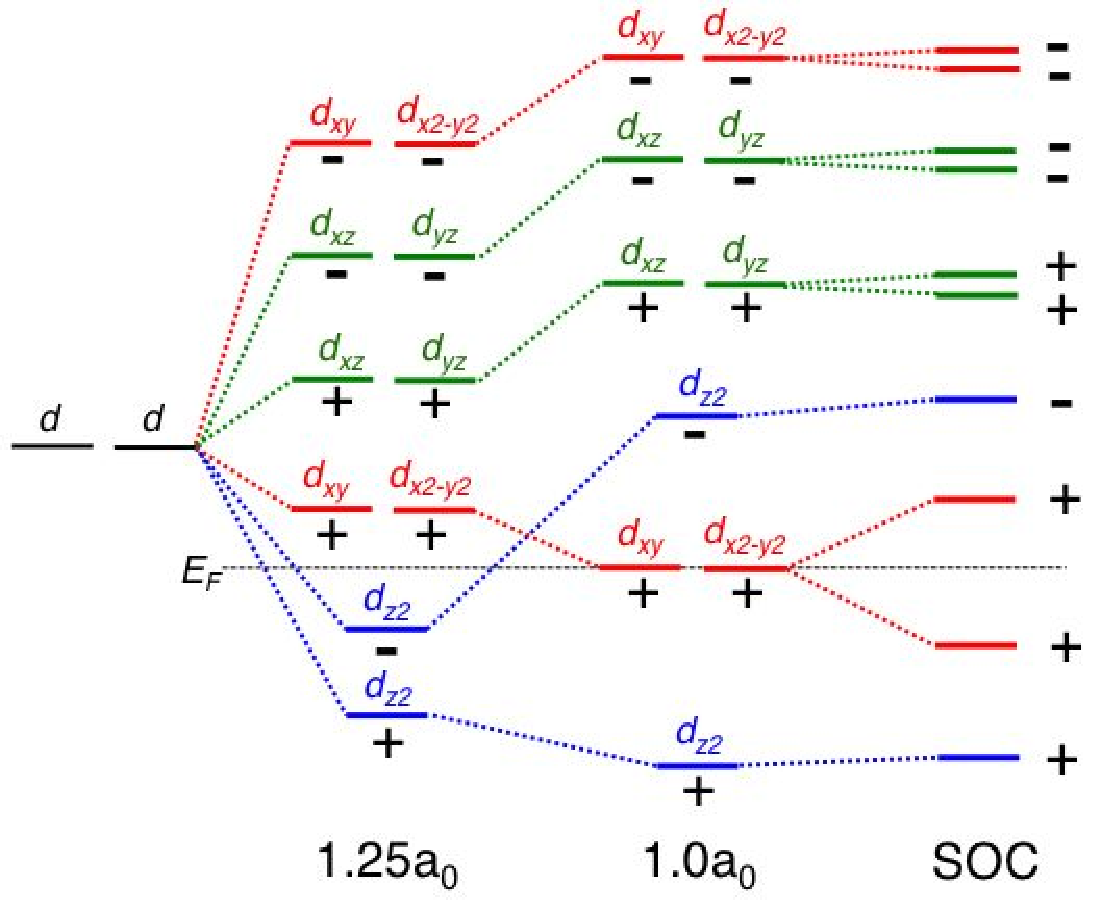}
\end{center}
\caption{(Color online) 
A schematically illustration of the evolution of the electronic bands at the $\Gamma$ 
point as the lattice constant is uniformly decreased to the optimized one $a_0$ 
without the SOC. The SOC is incorporated only in the left most figure.  
The bands are mainly drawn for W d orbitals, which have the highest contribution near the Fermi states.
W d-orbitals are split by crystal fields, chemical bondings, and SOC. $a_0$ is the optimized lattice constant and $E_f$ is the Fermi energy.} 
\label{fig:FIG6}
\end{figure}

The evolution of the electronic bands at the $\Gamma$ point is summarized schematically in Fig.~\ref{fig:FIG6}. 
When the lattice constant is very large ($\agt1.2a_0$, where $a_0$ is the optimized lattice constant in 
Sec.~\ref{sec:structure}) and the SOC is absent, W$_2$HfC$_2$O$_2$ is a trivial band insulator in 
which the two topmost valence bands are made of bonding and non-bonding 
states of $d_{3z^2-r^2}$ orbitals with even and odd parities, respectively, 
while the lowest conduction bands are doubly degenerated and formed by a bonding state of 
$d_{xy}$/$d_{x^2-y^2}$ orbitals with even parity. 
As the lattice constant is reduced, the hybridization between neighboring $d$ orbitals increases. 
As a result, the bonding state of doubly degenerate $d_{xy}$/$d_{x^2-y^2}$ orbitals shift downward below 
the Fermi energy, while the non-bonding state of $d_{3z^2-r^2}$ orbital moves above 
the Fermi energy. Since these two states have opposite parities, this is precisely when the band inversion 
occurs. Note that the Fermi energy remains exactly at the doubly degenerate $d_{xy}$/$d_{x^2-y^2}$ states 
once the band inversion occurs, and therefore the system remains semi-metallic. 
However, as soon as the SOC is introduced, the doubly degeneracy is lifted and the system becomes a 
nontrivial insulator. 
The role of the SOC is thus to induce a finite band gap but not a band inversion, 
which has similarly been observed in ZrTe$_5$~\cite{weng_2014}, square-octagonal 
WS$_2$~\cite{sun2015}, MX (M=Zr, Hf; X= Cl, Br, I)~\cite{zhou2015}, and W$_2$CO$_2$ systems~\cite{weng2015_2}.

It is worth summurizing the differences and similarities between the current study with 
our previous study in Ref.~\cite{weng2015_2}. Mo$_2$CO$_2$ and W$_2$CO$_2$ studied in 
Ref.~\cite{weng2015_2} belong to the family of functionalized M$_2$X MXenes with two transition metal 
layers, while the systems studied here such as Mo$_2$TiC$_2$ and Mo$_2$ZrC$_2$ belong to the family of 
ordered double transition metals carbides M$'_2$M$''$C$_2$ MXenes with three transition metal layers. 
For Mo$_2$CO$_2$ and W$_2$CO$_2$ as well as M$'_2$M$''$C$_2$O$_2$ 
(M$'$= Mo, W; M$''$= Ti, Zr, Hf), the SOC is essential to open a finite band gap, but does not induce the band inversion. The mechanism of the band inversion in M$'_2$M$''$C$_2$O$_2$ MXenes described above is 
similar to that in Mo$_2$CO$_2$ and W$_2$CO$_2$.The resulting topological band gap in the M$'_2$M$''$C$_2$ MXenes 
can be as large as that in W$_2$CO$_2$ (0.19 eV within the GGA and 
0.47 eV within the hybrid functional).

We have also studied the electronic structures of M$'_2$M$''_2$C$_3$O$_2$ (M$'$= Mo, W; M$''$= Ti, Zr, Hf), 
i.e., MXenes with four transition metal layers. 
Similar to the M$'_2$M$''$C$_2$O$_2$ systems, they have
hexagonal structures with $P\bar{3}m1$ space group symmetry and the BB model represents the 
most stable atomic configuration. 
The total energies, crystal structures, and band structures without and with the SOC are
shown in Supplemental Material~\cite{supplementary}. 
From the band structures and the Z$_2$ index calculations, we find that the M$'_2$M$''_2$C$_3$O$_2$ 
MXenes are nontrivial topological semimetals. 
The orbital characters of the bands near the Fermi energy around the $\Gamma$ point are very similar to those in 
the M$'_2$M$''_2$C$_2$O$_2$ systems. 
Because of the local $D_{3d}$ hexagonal symmetry, 
the touching valence and conduction bands at the $\Gamma$ point are mainly dominated by 
$d_{xy}$ and $d_{x^2-y^2}$ orbitals of M$'$ and M$''$. 
In Mo$_2$Hf$_2$C$_3$O$_2$ and W$_2$Hf$_2$C$_3$O$_2$, the band inversion is induced by the 
SOC at the $\Gamma$ point between W $d_{xy}/d_{x^2-y^2}$ and W $d_{3z^2-r^2}$ bands, 
while in the other  M$'_2$M$''_2$C$_3$O$_2$  (M$'$ =Mo, W; M$''$=Ti, Zr) MXenes,
the SOC only induces the band gap 
at the $\Gamma$ point but the band inversion occurs due to the hybridization of 
the neighboring orbitals, similar to the M$'_2$M$''$C$_2$O$_2$ systems.

\section{CONCLUSION}

Here, we have searched for new topological insulators in recently synthesized double transition metal 
carbide MXenes, and found that M$'_2$M$''$C$_2$O$_2$ (M$'$ = Mo, W; M$''$= Ti, Zr, Hf) are 
topological insulators with the largest band gap of 0.285 eV (0.401 eV within the hybrid functional) 
for W$_2$HfC$_2$O$_2$. 
The large band gap, resulting from the strong SOC in transition metals M$'$ and M$''$, is an attractive feature 
in these newly proposed topological insulators since 
the experimental realization of these systems with large band gaps would pave the way for practical applications 
of topological insulators at room temperature. In addition, we have found that
M$'_2$M$''_2$C$_3$O$_2$ with four transition metal layers are topological semimetals.
Owing to various compositional and surface functional possibility, as well 
as thickness control, we expect more topological insulators and topological semimetals can be found 
in the MXene family.

\acknowledgments 

We gratefully thank Dr. Yan Sun for his fruitful discussions and helpful comments and analyses. The calculations were performed on 
Numerical Materials Simulator at National Institute for Materials Science and RIKEN supercomputer system (HOKUSAI GreatWave).

\end{document}